\documentclass[aps,pre,superscriptaddress,showpacs,amsmath,amssymb,,twocolumn]{revtex4}

\usepackage{graphicx}
\usepackage{bm}% bold math
\usepackage{color}
\usepackage{amsmath}

\def\Vect#1{\mbox{\boldmath $#1$}}

\def\phij{\phi_{J}}
\def\sS{\sigma_{\rm y}}
\def\sa{\sigma_{\rm a}}
\begin{document}

\title{Avalanche contribution to shear modulus of granular materials}

\author{Michio Otsuki}
\email[]{otsuki@riko.shimane-u.ac.jp}
\affiliation{Department of Materials Science, Shimane University, Matsue 690-8504, Japan}
\author{Hisao Hayakawa }
\affiliation{Yukawa Institute for Theoretical Physics, Kyoto University, Kyoto 606-8502, Japan}

%\publishedin{%         %Write this ONLY in cases of addenda and errata
%Prog.~Theor.~Phys.\ \textbf{XX} (19YY), page.}

%\recdate{Mmmmm DD, YYYY}%            %Editorial Office will fill in this.

\begin{abstract}
Shear modulus of frictionless granular materials 
near the jamming transition
under oscillatory shear is numerically investigated.
It is found that the shear modulus $G$ satisfies a scaling law
to interpolate between $G \sim (\phi - \phij)^{1/2}$ and 
$G \sim \gamma_0^{-1/2}(\phi-\phi_J)$ 
for a linear spring model of the elastic 
interaction between contacting grains, where $\phi$, $\phij$, and $\gamma_0$ 
are, respectively, the volume fraction of grains, the fraction at the jamming point, and the amplitude of the oscillatory shear.
The linear relation between the shear modulus and 
$\phi - \phij$ can be understood 
by slip avalanches.
\end{abstract}
\date{\today}


\maketitle

\section{Introduction}

% Jamming
Amorphous materials consisting of densely packed
particles such as granular materials \cite{Jaeger},
colloidal suspensions \cite{Pusey}, 
emulsions, and foams \cite{Durian} have rigidity above a critical density,
while they lose rigidity below the critical density.
Such rigidity transition, known as the jamming transition,
has attracted much attention among researchers in these days \cite{Liu}.

% critical behavior frictionless particles
In the vicinity of the jamming point
taking place at the volume fraction of the jamming point $\phi_J$,
amorphous materials exhibit critical behavior.
Assemblies of frictionless particles
exhibit a mixed transition, in which
the coordination number shows a discontinuous transition,
while the pressure, the elastic moduli, and the characteristic frequency
of the density of state exhibit continuous transition
\cite{OHern02, OHern03,Wyart}.
Moreover, critical scaling laws, similar to those
observed in equilibrium critical phenomena,
exist in the rheology of the sheared disordered particles 
\cite{Majmudar,Olsson,Hatano07,Hatano08,Tighe,Hatano10,Otsuki08,Otsuki09,Otsuki10,Nordstrom,Olsson11,Vagberg,Otsuki12,Ikeda,Olsson12}.
On the other hand,
assemblies of soft frictional grains exhibit
a discontinuous transition associated with
a hysteresis loop and a discontinuous shear-thickening
in the rheology under steady shear
\cite{Otsuki11,bob_nature,Chialvo,Brown,Seto,Fernandez,Heussinger,Bandi,Wyart14}.

\begin{figure}
  \includegraphics[width=\linewidth]{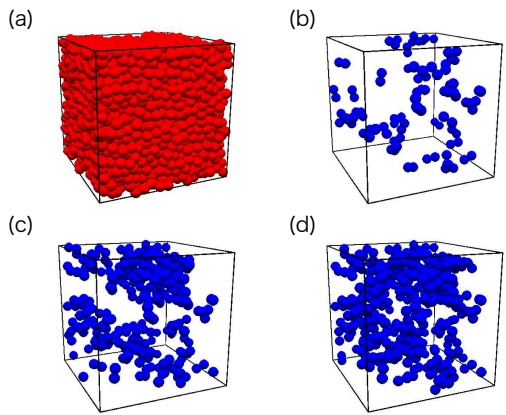}
  \caption{(Color online) Time evolution of a frictionless granular system 
of a linear spring between contacting grains for the packing fraction
  $\phi = 0.660$ under an oscillatory shear,
  where the jamming fraction $\phij$ is $0.6494$. 
  (a) A snapshot of grains without shear strain, i.e. $\gamma = 0$.
  (b) A snapshot of the grains whose bonds between contacting pairs
  at $\gamma = 0$ are broken at $\gamma = 1.2 \times 10^{-4}$.
  (c) A snapshot of the grains whose bonds between contacting pairs
  at $\gamma = 0$ are broken at $\gamma = 4.8 \times 10^{-4}$.
  (d) A snapshot of the grains whose bonds between contacting pairs
  at $\gamma = 0$ are broken at $\gamma = 7.5 \times 10^{-4}$.}
   \label{conf}
\end{figure}

% Shear modulus
The shear modulus $G$,
the ratio of the shear stress to the
shear strain,
is one of the most important quantities 
to characterize the jamming transition.
It is well known that
$G$ slightly above the jamming point satisfies the scaling
\begin{equation}
G \sim (\phi - \phij)^{1/2}
\label{Gsqrt}
\end{equation}
for grains interacting by a linear spring model,
  where $\phi$ is the volume fraction
\cite{OHern02, OHern03,Wyart}.
This power law 
as well as the frequency dependence of $G$ 
can be explained by the analysis
of the soft mode, 
and the validity of these laws are verified through simulations 
\cite{Wyart05,Tighe11}.
On the other hand, Refs. \cite{Mason, Yoshino} have recently reported that
$G$ might obey a different power law of the excess volume fraction,
$\phi - \phij$, as
\begin{equation}
G \sim \gamma_0^{-c} (\phi - \phi_J)
\label{Gprop}
\end{equation}
through an experiment and a simulation of soft
spherical particles at finite temperature
though the strain amplitude $\gamma_0$ dependence with 
an exponent $c$ has not been discussed.
The conflict between Eqs. (1) and (2) may be understood 
from the amplitude of the shear strain.
Indeed,
conventional studies assume that 
the contact network 
is unchanged during the process
because of an infinitesimal amplitude of the shear strain,
but might be inappropriate 
for a finite strain even near the jamming point.
In fact, as shown in Fig. \ref{conf}
obtained from a simulation under an oscillatory shear, 
many bonds between contacting grains
near the jamming point 
are broken
under the shear strain $\gamma$ larger than $10^{-4}$,
which causes slip avalanches distributed 
in a broad range of sizes \cite{Dahmen98,Dahmen11,Dobrinevski}.
To interpolate previously reported relations, Eqs. \eqref{Gsqrt} and 
\eqref{Gprop},
we postulate the scaling for the shear modulus:
\begin{equation}
G(\phi,\gamma_0) = G_0 (\phi - \phij)^a 
{\mathcal G} \left ( \gamma_0  / (\phi - \phij)^{b} \right ),
\label{G_scale:eq}
\end{equation}
where $a$ and $b$ are the critical exponents,
and $G_0$ is the characteristic shear modulus,
which is determined from the elasticity and the diameter of grains.
We also assume that the scaling function ${\mathcal G}(x)$
satisfies
\begin{equation}
\lim_{x\to 0}{\mathcal G}(x)  =  \mbox{const.}, \quad 
\lim_{x\to \infty}{\mathcal G}(x)  =  x^{-c}.
\label{G:eq1}
\end{equation}
To be consistent with the known results, the exponents
should satisfy $a=1/2$ and $a+bc=1$.
A similar analysis on the non-linear rheology 
of an unchanged contact network is reported
in Ref. \cite{Wyart08},
but they do not discuss the effect of the slip avalanches.
It should be noted that the plastic-elastic rheology
of jammed granular materials under large strain amplitude
is studied
in Ref. \cite{Otsuki13},
but the studies of the shear modulus depending 
on the stress avalanche by the shear strain
do not exist as long as we know.

In this paper, we numerically study the behavior of the shear modulus $G$
of granular materials
near the jamming point $\phij$
under an oscillatory shear.
In Sec. \ref{Setup:sec}, we explain our setup and model.
In Sec. \ref{Numerical:sec}, 
we present the details of our numerical results.
In Sec.  \ref{Theory:sec}, we phenomenologically estimate the values of
the exponents $a$, $b$, and $c$ we have introduced.
 We determine the values of exponent $a$
 in Eq. \eqref{G_scale:eq} in terms of a phenomenological argument
in Sec. \ref{ab:sec}, 
estimate the exponent $c$ in the asymptotic form  \eqref{G:eq1}
caused from  the slip avalanches
in Sec. \ref{c:sec},
and discuss the exponent $b$
 in Eq. \eqref{G_scale:eq}
in Sec. \ref{b:sec}.
In Sec. \ref{DC:sec}, we discuss and conclude our results.
In Appendix \ref{phij:sec}, we explain the method to determine
the jamming transition point.
In Appendix \ref{Avalanche:sec}, 
we re-derive the size distribution
of the avalanche obtained in Ref. \cite{Dahmen98}.

\section{Setup of our simulation}
\label{Setup:sec}

%system

Let us consider a three-dimensional frictionless granular assembly
in a cubic box of the linear size $L$. 
The system includes $N$ spherical grains,
where each of them has an identical mass $m$.
The position and the velocity of the grain $i$ are, respectively, denoted by
$\Vect{r}_i$ and $\Vect{v}_i$.
There exist $4$ types of grains for diameter,
  $0.7 d_0$, $0.8 d_0$, $0.9 d_0$,
and $d_0$,
where number of each species is $N/4$.
Throughout this paper, we use the volume fraction 
$\phi$ to characterize the density of the grains.

Because the grains are frictionless,
the contact force has only the normal component of 
the elastic force  $\Vect{f}_{ij}^{\rm (el)}$
and the dissipative force  $\Vect{f}_{ij}^{\rm (dis)}$,
which are respectively given by
\begin{eqnarray}
\Vect{f}_{ij}^{\rm (el)} &= &k (d_{ij} - r_{ij})^\Delta
\Theta (d_{ij} - r_{ij}) \Vect{n}_{ij}, \label{Fel}\\
\Vect{f}_{ij}^{\rm (dis)} &= &-\eta v_{ij} 
\Theta (d_{ij} - r_{ij}) \Vect{n}_{ij} \label{Fdis}
\end{eqnarray}
with 
the elastic constant $k$,
the viscous constant $\eta$, 
the diameter $d_i$ of grain $i$,
$\Vect{r}_{ij} \equiv \Vect{r}_{i} - \Vect{r}_{j} $, 
$\Vect{n}_{ij} \equiv \Vect{r}_{ij}/{r}_{ij}$,
$r_{ij} \equiv |\Vect{r}_{ij}|$,
$d_{ij} \equiv (d_i + d_j)/2$, and 
$v_{ij} \equiv (\Vect{v}_{i}- \Vect{v}_{j}) \cdot \Vect{n}_{ij}$. 
Here, $\Theta(x)$ is the Heaviside step function satisfying $\Theta(x)=1$
for $x \ge 0$ and $\Theta(x)=0$ for otherwise.
The exponent $\Delta$ characterizes the elastic repulsive interaction,
i.e. $\Delta = 3/2$ for spheres of Hertzian contact force,
and $\Delta = 1$ for the linear spring model.
Note that 
the characteristic shear modulus $G_0$ 
introduced in Eq. \eqref{G_scale:eq} corresponds to $k d_0^{\Delta - 2}$.

In this paper, we apply an oscillatory shear  
along the  $y$ direction under the Lees-Edwards boundary condition \cite{Evans}.
As a result, there exists macroscopic displacement 
only along the $x$ direction.
The time evolution of such a system, known as the SLLOD system \cite{Evans}, is given by
\begin{eqnarray}
\frac{d \Vect{r}_i}{dt} & = & \frac{\Vect{p}_i}{m} + \dot \gamma (t) y_i \Vect{e}_x,
\label{SLLOD:1} \\
\frac{d \Vect{p}_i}{dt} & = & \sum_{j \neq i} \{ \Vect{f}^{\rm (el)}_{ij}+  \Vect{f}^{\rm (dis)}_{ij} \}- 
\dot \gamma (t) p_{i,y} \Vect{e}_x,
\label{SLLOD:2}
\end{eqnarray}
where $\Vect{p}_i$ and $\Vect{e}_x$ are respectively the peculiar momentum and the unit vector parallel to the $x$ direction.

We use the viscous constant $\eta = 1.0 \sqrt{mkd_0^{\Delta - 1}}$,
which corresponds to the constant restitution coefficient
$e = 0.043$ for $\Delta = 1$. We adopt the leapfrog algorithm, the second-order
accuracy in time with the time interval $\Delta t = 0.2 \tau$, 
where $\tau$ is 
the characteristic time of the stiffness, i.e. $\tau= \sqrt{md_0^{1-\Delta} /k}$.
The number $N$ of the particles is $16000$ except in
Appendix \ref{phij:sec}, where we estimate the jamming point $\phij$
from a finite size scaling.
We have verified that the shear modulus is almost independent of the system size
for $N\ge4000$.

\begin{figure}
  \includegraphics[width=\linewidth]{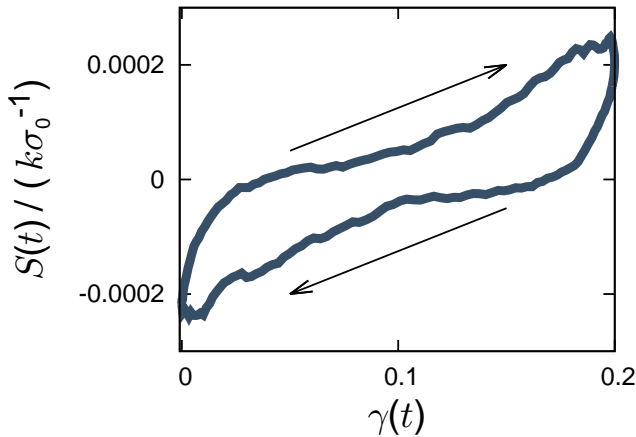}
  \caption{(Color online) The shear stress $S(t)$ against $\gamma(t)$ 
  with $\phi = 0.652$ and $\gamma_0 = 0.1$ for $\Delta = 1$.
  }
   \label{st_ga}
\end{figure}

We randomly place the grains in the system as an initial state,
and wait until the kinetic energy of each grain becomes
smaller than $10^{-14} kd_0^{1+\Delta}$.
Then, we apply the shear with the shear rate
\begin{eqnarray}
\dot \gamma (t) = \gamma_0 \omega \sin(\omega t),
\label{dgammat}
\end{eqnarray}
where  time $t$ is measured from the relaxed static configuration
and 
$\omega$ is the angular frequency of the oscillatory shear. 
From Eq. \eqref{dgammat},
the shear strain is given by
\begin{eqnarray}
\gamma (t) = \gamma_0 \left \{ 1 - \cos(\omega t) \right \}.
\label{gammat}
\end{eqnarray}
We examine the shear modulus for various strain amplitudes
$\gamma_0 = 10^0, 10^{-1}, 10^{-2}, 10^{-3}, 10^{-4},$ and $10^{-5}$
for $\omega \tau = 10^{-4}$ \cite{gamma:note}.
We analyze the real part of the complex shear modulus \cite{Doi}
(storage modulus) defined by
\begin{equation}
G(\phi,\gamma_0,\omega) = - \frac{\omega}{\pi} \int_{t_0(\gamma_0)}^{2\pi/\omega+t_0(\gamma_0)} dt \frac{S(t) \cos (\omega t)}{\gamma_0},
\label{G:def}
\end{equation}
where $t_0(\gamma_0)$ is the time when $\gamma(t) = 0$ 
under the strain amplitude $\gamma_0$.
Here, the shear stress $S(t)$ is calculated from
\begin{eqnarray}
S(t)  & = &  -\frac{1}{L^2}\left <    \sum_i^N \sum_{j>i} 
r_{ij,x}(t) \left \{ f_{ij,y}^{\rm (el)}(t) +f_{ij,y}^{\rm (dis)}(t)  \right \}
\right > \nonumber \\ 
& & -\frac{1}{L^2} \left <   \sum_{i=1}^N \frac{p_{x,i}(t)p_{y,i}(t)}{2m} \right >.
\label{S:calc}
\end{eqnarray}
In this paper, we do not analyze the loss modulus because
i) it has only the linear dependence on $\omega$ 
in our simulation and
ii) it seems to be independent of  density.
Note that the stress $S(t)$ exhibits a strong nonlinearity
on the strain $\gamma(t)$ as shown in Fig. \ref{st_ga},
where we plot the shear stress $S(t)$ against $\gamma(t)$ 
with $\phi = 0.652$ and $\gamma_0 = 0.1$ for $\Delta = 1$.
It should also be noted that 
$G$ is almost independent of $\omega$
for $\gamma_0 \le 10^{-2}$.
We, thus, investigate only $\gamma_0$ and $\phi$ dependence of $G$
in this paper.

\section{Numerical Results}
\label{Numerical:sec}

\begin{figure}
  \includegraphics[width=\linewidth]{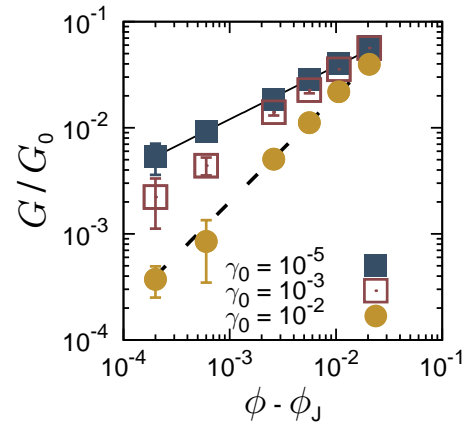}
  \caption{(Color online) 
The shear modulus $G$ against
the excess volume fraction $\phi - \phij$
with $\gamma_0 = 10^{-5}, 10^{-3}$,
and $10^{-2}$ for $\Delta = 1$.
The solid and the dashed lines represent the power law functions
with the exponent $1/2$ and $1$, respectively.
  } 
   \label{G_large}
\end{figure}

In Fig. \ref{G_large},
we plot $G$ against $\phi - \phij$ with $\gamma_0 = 10^{-5}, 10^{-3}$ and $10^{-2}$ for $\Delta = 1$.
It should be noted that the jamming point $\phij$ is numerically estimated
as $0.6494$ by the method explained in Appendix \ref{phij:sec}.
For the smallest strain amplitude ($\gamma_0 = 10^{-5}$), 
$G$ reproduces the well known behavior Eq. \eqref{Gsqrt}
\cite{OHern03,Wyart05},
but $G$ seems to satisfy
Eq. \eqref{Gprop}
for large $\gamma_0=10^{-2}$.
Thus, it is natural to postulate the scaling form 
Eq. \eqref{Gprop} to interpolate two equations.

%critical scaling

\begin{figure}
  \includegraphics[width=\linewidth]{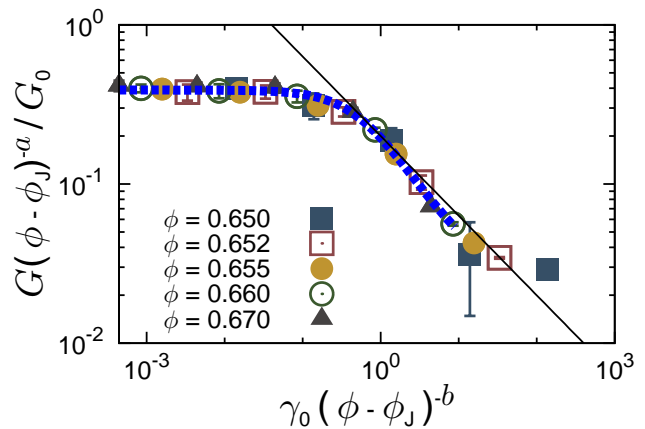}
  \caption{(Color online) Scaling plot of $G$ characterized by
  Eq.  \eqref{G_scale:eq} with $\phi = 0.650, 0.652, 0.655, 0.660, 0.670$ 
  and $\gamma_0 = 10^{-4}, 10^{-3}, 10^{-2}, 10^{-1}$ 
  for $\Delta = 1$. The dashed line is the scaling function given by
Eq. \eqref{G_fit}. 
The solid line represents the second equation
in Eq. \eqref{G:eq1}
with the exponent $c = 1/2$.
  } 
   \label{G_scale}
\end{figure}

Figure \ref{G_scale} shows the scaling plot based on Eq. \eqref{G_scale:eq}
for $\Delta = 1$.
This figure supports the scaling ansatz, Eq. \eqref{G_scale:eq},
where we have used exponents
\begin{equation}
a = 0.50 \pm 0.02, \quad b = 0.98 \pm 0.02.
\label{ab_est:eq}
\end{equation}
The exponents are determined
by the Levenberg-Marquardt algorithm \cite{NC},
where we use the functional form for the scaling function: 
\begin{equation}
{\mathcal G} (x) = \frac{B_0}{1 + e^{\sum_{n=1}^3 B_n (\log x)^n}}
\label{G_fit}
\end{equation}
with fitting parameters $B_0=0.39 \pm 0.03$, $B_1 = 1.1 \pm 0.06$, $B_2 = -0.08 \pm 0.04$, 
and $B_3 = -0.008 \pm 0.008$.
Here, we use the critical fraction estimated from a finite size scaling
in Appendix \ref{phij:sec}.
It should be noted that the estimated values of the exponents
do not change within the error margin
if we use $\phij$ as a free parameter
in the Levenberg-Marquardt algorithm. 
From Fig. \ref{G_scale},
the estimated exponent $c$ in Eq. \eqref{G:eq1} is approximately given by $1/2$.
From Eq. \eqref{ab_est:eq} and $c=1/2$,
we obtain $a + bc = 0.99 \pm 0.02$,
which also supports Eq. \eqref{Gprop}.

\begin{figure}
  \includegraphics[width=\linewidth]{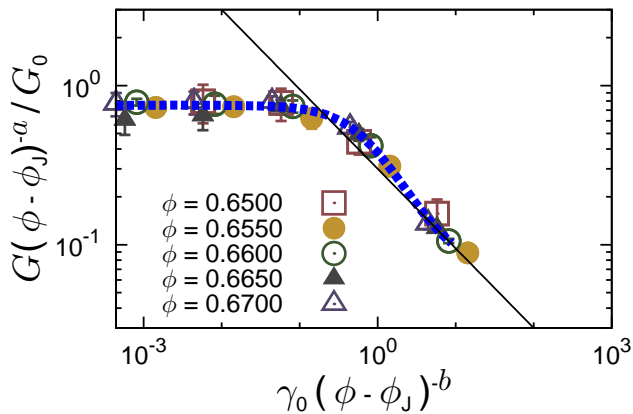}
  \caption{(Color online) Scaling plot of $G$ characterized by
  Eq.  \eqref{G_scale:eq} 
  with $\gamma_0 = 10^{-5}, 10^{-4}, 10^{-3}, 10^{-2}$ 
  and $\gamma_0 = 10^{-4}, 10^{-3}, 10^{-2}, 10^{-1}$ 
for $\Delta=3/2$. The dashed line is the scaling function given by
Eq. \eqref{G_fit}. 
The solid line represents the second equation
in Eq. \eqref{G:eq1}
with the exponent $c = 1/2$.
  } 
   \label{G_scale_Hertz}
\end{figure}

Figure \ref{G_scale_Hertz}
confirms the validity of Eq. \eqref{G_scale:eq}
for $\Delta = 3/2$,
  where the scaling exponents are numerically estimated as
\begin{equation}
a = 0.99 \pm 0.02, \quad b = 0.98 \pm 0.01
\label{exp:Hertz}
\end{equation}
with 
the fitting parameters $B_0 = 0.76 \pm 0.16, B_1 = 1.1 \pm 0.18, B_2 = -0.089\pm0.073,B_3=-0.020 \pm 0.016$ 
and
the critical fraction $\phij = 0.6486 \pm 0.0001$,
which is numerically estimated
by the method explained in Appendix \ref{phij:sec}.
The exponent $c$ in Eq. \eqref{G:eq1} is approximately given by $1/2$.
It should be noted that the estimated values of the exponents
do not change within the error margin
if we use $\phij$ as a free parameter.

\section{Phenomenological explanation}
\label{Theory:sec}

In this section,
we try to evaluate the exponents for the scaling law
Eqs. \eqref{G_scale:eq} and \eqref{G:eq1}
in terms of a mean-field like phenomenological argument.
In the first part, we derive the exponent $a$
in Eq. \eqref{G_scale:eq}.
In the second part, we determine the exponent $c$
in Eq. \eqref{G:eq1}.
Finally, we discuss the exponent $b$ in Eq. \eqref{G_scale:eq}.

\subsection{Exponents $a$}
\label{ab:sec}
%shear modulus for infinitesimal shear

Let us derive the exponent $a$.
The relationship between the mean-field 
compress force $f$ and the compression 
$\delta$, $f\sim k_{\rm eff} 
\delta \sim \delta^\Delta$
with the effective spring constant 
$k_{\rm eff}$, 
we immediately obtain the relation
$k_{\rm eff}\sim \delta^{\Delta-1}$.
In the vicinity of the jamming point,
the compression $\delta$ should satisfy
$\delta\sim \phi-\phi_J$ \cite{Otsuki08}.
On the other hand, from Refs. 
\cite{Wyart05,Tighe12},  we may deduce
$G/k_{eff}  \sim \delta z\sim  \sqrt{\phi-\phi_J}$,
where $\delta z$ is the excess 
coordination number 
and we have used the well known relation:
$\delta z\sim \sqrt{\phi-\phi_J}$ \cite{Wyart}.
Thanks to the above relations, 
we reach
 $G\sim (\phi-\phi_J)^{\Delta-1/2}$.
Therefore, we obtain the exponent $a$ as
\begin{eqnarray}
a = \Delta - 1/2.
\label{a_gen:eq}
\end{eqnarray}
Equation \eqref{a_gen:eq} is consistent with 
the numerical estimation given by Eqs. \eqref{ab_est:eq} and \eqref{exp:Hertz}
for $\Delta = 1$ and $3/2$, respectively.

\subsection{Exponent $c$}
\label{c:sec}
%generalized elastic-plastic model

%Let us evaluate  the exponent $c$ in Eq. \eqref{G:eq1}.
%For the linear spring model, 
%the exponent should be $c=1/2$ 
%because of $a+bc=1$, $a=1/2$, and $b=1$ in Eq. \eqref{b:eq}.  
%However, it seems that Eq. \eqref{Gprop} and 
%$a+bc=1$ cannot be used for the general $\Delta$. 
%Therefore, we should determine the exponent $c$ from an independent argument.

We assume
that the shear stress under the oscillatory shear
is described by a generalized elastic-plastic model \cite{Lubarda}.
%as illustrated in Fig. \ref{EP:fig} .
%\begin{figure}
%  \includegraphics[width=\linewidth]{EP.eps}
%  \caption{ A schematic picture of 
%a generalized elastic-plastic model
% consisting of infinite number of
%series connections with an elastic element of equal shear modulus $G_0$
%and a slip element of unequal yield stress $s$.
%  } 
%   \label{EP:fig}
%\end{figure}
Here, the elastic-plastic model consists of infinite number of
series connections with an elastic element of equal shear modulus $G_0$
and a slip element characterized by the stress drop $s$
in each avalanche process.
We assume that the time evolution of the shear stress $S(t)$ is given by
\begin{equation}
S(t) = \int_0^{\infty} \rho(s) \tilde S(s,t) \ ds,
\label{EP:eq}
\end{equation}
where $\tilde S(s,t)$ is the stress of an individual element 
having the stress drop $s$, and $\rho(s)$ is the probability density
of the stress drop.

\begin{figure}
  \includegraphics[width=\linewidth]{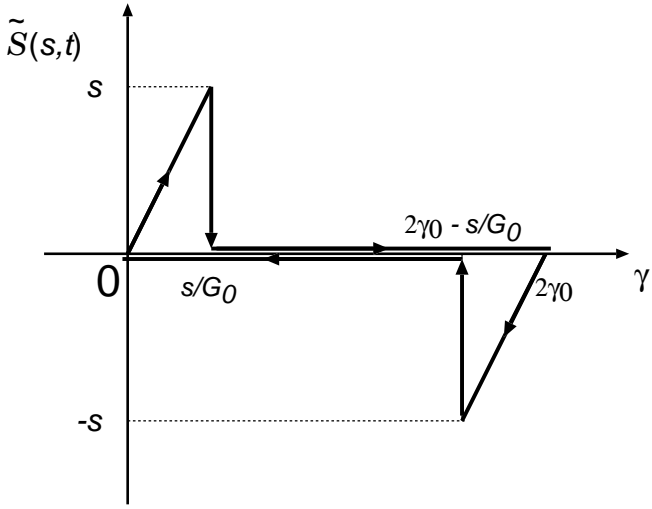}
  \caption{
The stress $\tilde S(s,t)$ of an individual element
for $0 \le t \le 2\pi/\omega$.
  } 
   \label{EPI}
\end{figure}

We assume that the individual stress $\tilde S(s,t)$
for $0 \le t \le 2\pi/\omega$ 
behaves as a linear function of the strain $\gamma(t)$
given by Eq. \eqref{gammat}
until $|\tilde S(s,t)|$ reaches the maximum value $s$,
while it drops to $0$ when $|\tilde S(s,t)|$
  exceeds $s$ due to the breakdown of the contact network.
Thus, $\tilde S(s,t)$ satisfies
\begin{eqnarray}
\tilde S(s,t) = 
\left\{
  \begin{array}{ll}
  G_0 \ \gamma(t) & (0 \le \theta(t) < \theta_c) \\
  0 &  (\theta_c \le \theta(t) < \pi)\\
  G_0 \ (\gamma(t) - 2\gamma_0) & (\pi \le \theta(t) < \pi + \theta_c) \\
  0 &  (\pi + \theta_c \le \theta(t) < 2\pi),
  \end{array}
  \right.
  \label{sigmat}
\end{eqnarray}
as illustrated in Fig. \ref{EPI},
where
$\theta(t)$ is the phase of the shear strain:
\begin{equation}
\theta(t) = \omega t.
\label{theta}
\end{equation}
The explicit expression of the critical phase $\theta_c$ for $S(s,t) = s$
is given by
\begin{eqnarray}
\theta_c\left({s}/({G_0\gamma_0}) \right )  = \cos^{-1}\left (1-\frac{s}{G_0 \gamma_0} \right ),
\end{eqnarray}
where we have used 
Eqs. \eqref{gammat}, \eqref{sigmat}, and \eqref{theta}.

%avalanche size
The expression of the stress-strain relation
\eqref{EP:eq}
depends on 
the probability density $\rho(s)$,
which is predicted to satisfy 
\begin{eqnarray}
\rho(s) & = &
  A(\phi) s^{-3/2} e^{ -s / s_{\rm c}(\phi)}
  \label{rhos}
\end{eqnarray}
for $s_0(\phi) \le s$,
where $s_0(\phi)$ and $s_{\rm c}(\phi)$
are the lower cutoff and the characteristic stress drop, 
respectively
\cite{Dahmen98,Dahmen11,Dobrinevski}.
(The derivation and the numerical result for
$\rho( s)$ are presented in Appendix \ref{Avalanche:sec}.)
It should be noted that the cutoff size of the stress drop distribution should exist, because the rearrangement of one grain gives the minimum size of stress drop, though the cutoff might differ from $s_0(\phi)$. Here, we simply assume that the distribution lower than $s_0(\phi)$ does not contribute to the shear modulus.
Here, $A(\phi)$ is the normalization constant satisfying
$A(\phi)=1/\int_{s_0(\phi)}^{\infty} ds s^{-3/2} e^{ -s / s_{\rm c}(\phi)}$,
which depends on the volume fraction.

%calculation of G

Substituting Eq. \eqref{EP:eq} into Eq. \eqref{G:def},
we obtain 
\begin{eqnarray}
G = \int_0^\infty ds \tilde G(\gamma_0,s) \rho(s),
\label{G2:eq}
\end{eqnarray}
where $\tilde G(\gamma_0,s)$ is the shear modulus of the individual
element:
\begin{equation}
\tilde G(\gamma_0, s) = - \frac{\omega}{\pi} \int_0^{2\pi/\omega} dt \frac{\tilde S(s, t) \cos (\omega t)}{\gamma_0}.
\label{Gt:def}
\end{equation}
Substituting Eq. \eqref{sigmat} into Eq. \eqref{Gt:def},
we obtain
\begin{eqnarray}
\tilde G(\gamma_0, s) = G_0 F\left ( \frac{s}{G_0 \gamma_0} \right ),
\label{Gt2:eq}
\end{eqnarray}
where
\begin{eqnarray}
F(x) =
\left\{
  \begin{array}{ll}
  1  & (x \ge 1)  \\
   T(x) / \pi &   (x < 1)
  \end{array}
  \right.
\end{eqnarray}
with 
\begin{equation}
T(x) = \theta_c(x) -2\sin \theta_c(x) + 
  \frac{\sin 2\theta_c(x)}{2}.
\end{equation}
Substituting Eqs. \eqref{rhos} and \eqref{Gt2:eq}
into Eq. \eqref{G2:eq},
we obtain
\begin{eqnarray}
G = A(\phi) G_0 \int_{s_0}^{\infty} ds \
s^{-3/2} e^{ -s / s_{\rm c}(\phi)} F\left ( \frac{s}{G_0 \gamma_0} \right ).
\label{G3:eq}
\end{eqnarray}

Using $x = {s}/{G_0 \gamma_0}$
and expansion
$e^{- x (G_0\gamma_0)/s_{\rm c}} = 1 - x (G_0\gamma_0)/s_{\rm c} + \cdots$
for $s \gg s_{\rm c}(\phi)$,
we obtain
\begin{eqnarray}
G & = & A(\phi)G_0^{1/2}\gamma_0^{-1/2} \nonumber \\
& & \times \left \{ \int_{\frac{s_0}{G_0 \gamma_0}}^{\infty} dx \
x^{-3/2} F\left ( x \right ) + O\left(\frac{G_0\gamma_0}{s_{\rm c}}\right) \right \}.
\end{eqnarray}
For $s_0/G_0 \ll  \gamma_0 \ll s_{\rm c}/G_0$,
the second term in this equation is negligible and the lower cutoff of the integral can be
${s_0}/{(G_0 \gamma_0)} \to 0$.
Then, $G$ is approximately given by
\begin{eqnarray}
G \simeq A(\phi)G_0^{1/2}\gamma_0^{-1/2} \int_{0}^{\infty} dx \
x^{-3/2} F\left ( x \right ).
\label{G4:eq}
\end{eqnarray}
Because the integral in Eq. \eqref{G4:eq} is apparently converged,
we obtain
\begin{eqnarray}
c = 1/2
\label{c:eq}
\end{eqnarray}
from Eqs. \eqref{G_scale:eq} and \eqref{G:eq1}.

\subsection{Exponents $b$}
\label{b:sec}

From Eqs.  \eqref{a_gen:eq} and \eqref{c:eq} with
the aid of Eqs. \eqref{G_scale:eq} and \eqref{G:eq1},
Eq. \eqref{Gprop} for the general $\Delta$ is replaced by
\begin{equation}
G \sim \gamma_0^{-1/2} (\phi-\phi_J)^{\Delta + (b-1)/2}.
\end{equation}

It should be noted that the dimensions of the shear modulus $G$ 
and the pressure $P$ are identical, 
$G$
under large strain amplitude $\gamma_0$
might obey the same dependence on $(\phi-\phi_J)$
as that of $P$ \cite{OHern03}, $P \sim (\phi-\phi_J)^\Delta$,
which leads to $b=1$.
This is consistent with
the numerical estimation given by Eqs. \eqref{ab_est:eq} and \eqref{exp:Hertz}
for $\Delta = 1$ and $3/2$, respectively.
Thus, we believe that $b=1$ can be used in our setup,
which is consistent with the resect experiments
\cite{Coulais}.

\section{Discussion and Conclusion}
\label{DC:sec}
This section consists of two parts.
In the first part, we discuss our results,
and we conclude our work in the second part.

\subsection{Discussion}
\label{Discuss:sec}

Now, let us discuss our results.
First, we discuss the relationship between our result and the scaling law 
of $G$
proposed in Ref. \cite{Tighe11}.
Second, we compare our results with those on the power spectrum of 
the shear stress.
Finally, we mention the effect of the friction 
on the scaling for the shear modulus.

%the result of Tighe

Tighe reported that 
the shear modulus $G$ satisfies
a power law of the angular frequency $\omega$ for the oscillatory shear
at the jamming point:
\begin{equation}
G \sim \omega^{1/2}
\label{G:tighe}
\end{equation}
for an analysis of a model of emulsions \cite{Tighe11}.
In contrast, both our simulation and phenomenology 
suggest that the shear modulus is independent of $\omega$.
We believe that his viscous force preventing grains from the rotation and the sliding
is the origin of the nontrivial
relation \eqref{G:tighe} \cite{Tighe12},
which is not involved in our model
in Eq. \eqref{Fdis}.
This is the reason for
the absence of the $\omega$-dependence of $G$
in our results.

In a simulation and an experiment
of granular materials under steady shear \cite{Dahmen11,Dalton},
the power spectrum of the shear stress exhibits a non-trivial power law dependence on the frequency
$\omega$.
In contrast,
such a dependence of $G$ does not exist in our simulation
under oscillatory shear.
It should be noted that the power spectrum is directly related to the time 
correlation of the stress,
but the shear modulus $G$ is related to the average of the stress,
which is the origin of the different $\omega$-dependences.
To study power spectrum of the shear stress would be one of our future subjects.

%From Eq. \eqref{Gsqrt} and $\gamma_c(\phi) \sim \phi - \phij$,
%the characteristic stress $S_c(\phi)=G(\phi)\gamma_c(\phi)$
%for the appearance of the nonlinear elasticity
%is expected to be proportional to $(\phi - \phij)^{3/2}$
%for the linear spring model,
%which seems to be contradicted with the result
%of the dynamic yield stress $S_y(\phi) \propto \phi - \phij$ under steady shear
%\cite{Otsuki08,Otsuki09}.
%However, $S_y(\phi)$ is the shear stress at the steady state
%with infinitesimal shear rate $\dot \gamma$ but finite shear strain
% because of the relation $\gamma=\lim_{t\to \infty} \dot\gamma t$,
%and is not directly related to $S_c(\phi)$ for small shear strain.
%This is the reason for the different scalings of $S_c(\phi)$ and $S_y(\phi)$.

%effect of friction between particles

It is known that
the rheology is drastically affected by
friction between particles, at least, 
for assemblies of soft grains under steady shear
\cite{Otsuki11,Chialvo,Brown,Seto,Fernandez,Heussinger}.
The friction plays a key role to cause the shear thickening 
in rheology, and thus, study on the rheology of frictional grains under an oscillatory shear is practically important.
The friction dependence of the scaling law \eqref{G_scale:eq}
will be discussed elsewhere.

\subsection{Summary}
\label{Summary:sec}

%conclusion
In conclusion, we numerically study the frictionless
granular particles and propose a new scaling
law which interpolate
between $G \sim (\phi - \phij)^{\Delta - 1/2}$ 
for infinitesimal strain and 
$G \sim \gamma_0^{-1/2}(\phi-\phi_J)^{\Delta}$
for finite strain, where $\Delta$ is the exponent 
to characterize the local elastic interaction between contacting grains.
These scaling exponents are verified through our simulation. 
The scaling of the shear strain under the large strain 
can be understood by the theory of slip avalanches.

\begin{acknowledgments}
The authors thanks B. P. Tighe, K. Kamrin, H. Yoshino, K. Miyazaki,
S. Titel, and T. Yamaguchi
for fruitful discussions, and K. Saitoh and K. Suzuki for their
  critical reading of the manuscript. The authors also wish to thank
  Aspen Center for Physics, where parts of this work is developed. This
  work was supported by JSPS KAKENHI (Grant Nos. 25287098,
  22540398, and 25800220) and in part by the Yukawa International
  Program for Quark-Hadron Sciences (YIPQS). One of the authors (MO)
  appreciates the warm hospitality of Yukawa Institute for Theoretical
  Physics at Kyoto University and the discussions during the YITP
  workshop YITP-W-13-04 on ``Physics of glassy and granular materials",
  YITP-T-13- 03 on ``Physics of Granular Flow", and YITP-W-10-20 on
  ``Recent Progress in Physics of Dissipative Particles" to complete this
  work.
\end{acknowledgments}

\appendix

\section{Determination of transition point}
\label{phij:sec}

\begin{figure}[htbp]
  \includegraphics[width=\linewidth]{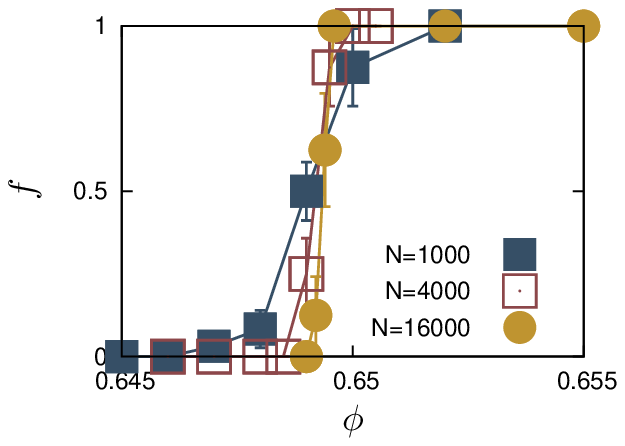}
  \caption{(Color online) The jammed fraction $f$ against $\phi$
  with different system sizes $N = 1000, 4000$, and $16000$
  for $\Delta = 1$.
  } 
   \label{f}
\end{figure}

\begin{figure}[htbp]
  \includegraphics[width=\linewidth]{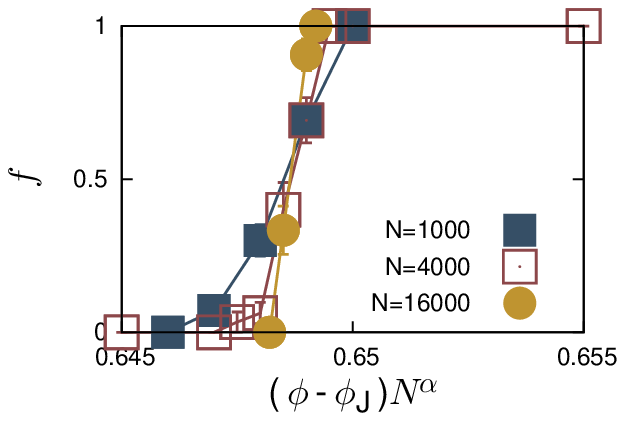}
  \caption{(Color online) The jammed fraction $f$ against $\phi$
  with different system sizes $N = 1000, 4000$, and $16000$
  for $\Delta = 3/2$.
  } 
   \label{f_Hertz}
\end{figure}

In this appendix, we explain how to determine the critical volume fraction $\phij$.
Here, we assume that $\phij$ is the volume fraction
where the pressure $P$ in the system of $N \to \infty$
becomes finite
under sufficiently small and slow shear strain.
We, thus, introduce $f$ as the fraction of samples where $P$
is larger than a threshold value $P_{\rm th} = 10^{-6} k d_0^{\Delta-2}$
for $\gamma_0 = 10^{-4}$ and 
  $\omega \tau = 10^{-4}$.
It should be noted that the estimated $\phij$
is independent of the choice of $P_{\rm th}$ 
within the error margin, at least,
for $ 5.0 \times 10^{-7} < P_{\rm th} / (k d_0^{\Delta-2}) < 1.0 \times 10^{-5}$ .
Here, $P$ is given by
\begin{eqnarray}
P  & = &   \frac{1}{3L^2}\left <    \sum_i^N \sum_{j>i} 
\Vect{r}_{ij} \cdot (\Vect{f}_{ij}^{\rm (el)} +\Vect{f}_{ij}^{\rm (dis)} ) 
\right >  \nonumber \\
& &
+
\frac{1}{3L^2} \left <   \sum_{i=1}^N \frac{|\Vect{p}_{i}|^2}{2m} \right >.
\label{P:calc}
\end{eqnarray}
Figures \ref{f} and \ref{f_Hertz} plot the jammed fraction $f$ against $\phi$
for $\Delta = 1$ and $3/2$, respectively.
Here, $f$ is zero for low $\phi$ and $f$ is finite 
for large $\phi$.
It should be noted that the slope of $f$ around $\phi = 0.65$ becomes steeper
as the system size increases.

\begin{figure}
  \includegraphics[width=\linewidth]{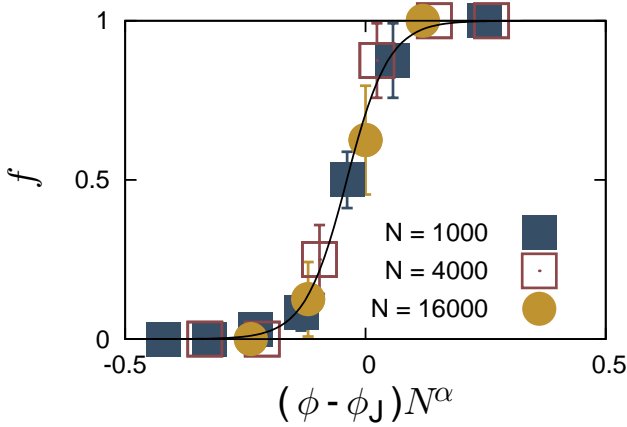}
  \caption{(Color online) Scaling plot of the jammed fraction $f$ 
  characterized by Eq. \eqref{f_scale:eq} for $\Delta = 1$. The solid line is the scaling
  function given by Eq. \eqref{F:eq}.
  } 
   \label{f_scale}
\end{figure}

\begin{figure}
  \includegraphics[width=\linewidth]{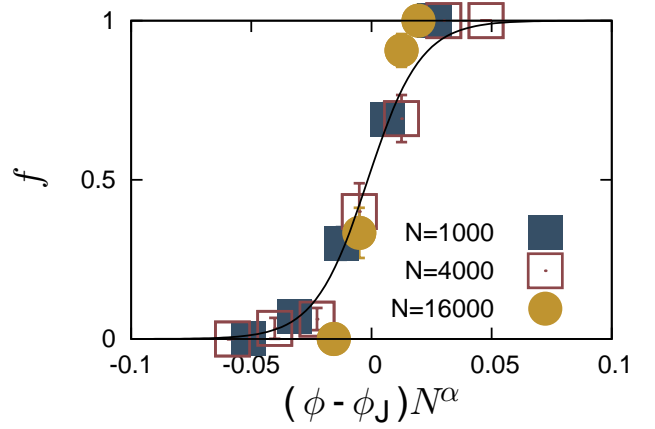}
  \caption{(Color online) Scaling plot of the jammed fraction $f$ 
  characterized by Eq. \eqref{f_scale:eq} for $\Delta = 3/2$. The solid line is the scaling
  function given by Eq. \eqref{F:eq}.
  } 
   \label{f_scale_Hertz}
\end{figure}

In order to determine $\phij$ from the data in Figs. \ref{f} and \ref{f_Hertz},
we assume $f(\phi, N)$ satisfies a scaling relation
\begin{equation}
f(\phi, N) = H((\phi - \phij)N^\alpha)
\label{f_scale:eq}
\end{equation}
with an exponent $\alpha$ and a scaling function $H(x)$
which satisfies $\lim_{x \to \infty} H(x) = 1$
and  $\lim_{x \to - \infty} H(x) = 0$.
Figures \ref{f_scale} and \ref{f_scale_Hertz} verify 
the assumption \eqref{f_scale:eq},
and thus, we can determine $\phij = 0.6494 \pm 0.0001$ and $\phij = 0.6486 \pm 0.0001$, respectively.
Here, we have assumed the functional form of the scaling function
as
\begin{equation}
H(x) = \left \{  1 + \tanh \left ( A_0  + A_1 x \right )\right \} / 2
\label{F:eq}
\end{equation}
with the fitting parameters $A_0 = 0.4 \pm 0.2, A_1 = 11 \pm 6$
and $\alpha = 0.66 \pm 0.07$ for $\Delta = 1$,
while $A_0 = 0.06 \pm 0.1, A_1 = 45 \pm 41$
and $\alpha = 0.42 \pm 0.12$ for $\Delta = 3/2$.

It should be noted that we estimate $\phij$
from $P$ at the state with $\gamma_0 = 10^{-4}$,
which is obtained by varying
the strain amplitude from $10^0$ to sequentially
decreasing values as $\gamma_0 = 10^0, 10^{-1}, 10^{-3}, 10^{-4}$.
The estimated value of $\phij$ depends on the detail of the protocol to 
decrease $\gamma_0$,
which might be the origin of the difference of $\phij$
for $\Delta = 1$ and $3/2$.

\section{Distribution of Avalanche size}
\label{Avalanche:sec}

In this appendix, we re-derive the
probability density $\rho(s)$ of the stress drop $s$
obtained in Refs. \cite{Dahmen98,Dahmen11}.

\subsection{Setup}

\begin{figure}[htbp]
  \includegraphics[width=0.7\linewidth]{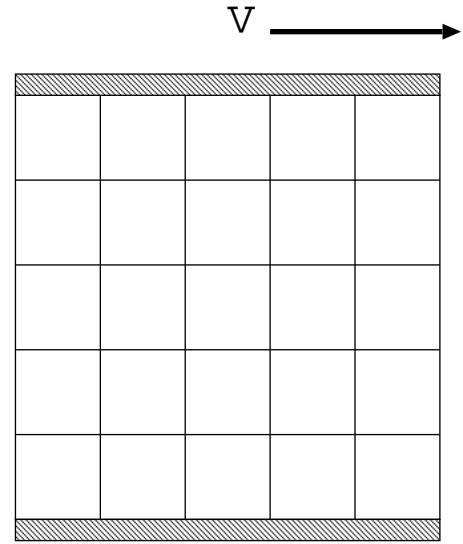}
  \caption{Illustration of the sheared model lattice.}
  \label{DModel}
\end{figure}

In Refs. \cite{Dahmen98,Dahmen11}, sheared granular materials
are modeled as a simplified lattice system on a coarse-grained scale
(larger than the grain diameter) consisting of $N'$ sites
and the linear size $L$.
We apply a strain by moving one boundary
at a slow speed $V$ (see Fig. \ref{DModel}).

In this setup, the local shear stress  $\sigma_i$ at site $i$
under the mean field approximation
may be given by
\begin{equation}
\sigma_i = K(Vt - u_i) + \frac{J}{N'} \sum_{j=1}^{N'} (u_j - u_i),
\label{si}
\end{equation}
where $u_i$ is the displacement at site $i$.
The first term on the right hand side (RHS) of Eq. \eqref{si}
represents the contribution of the global shear 
under the elastic constant $K$, which may satisfy the relation
$K \sim G_0/L$.
The second term on RHS of Eq. \eqref{si} represents the 
mean-field interaction
with the coupling constant $J/N'$.
We can rewrite Eq. \eqref{si} as
\begin{equation}
\sigma _i = KVt + J \bar u - (K+J) u_i,
\label{si_mo}
\end{equation}
where we have introduced
\begin{equation}
\bar u = \sum_{j=1}^{N'} u_j /N'.
\end{equation}
The stress $\sigma$ of the system is defined as the average of $\sigma_i$:
\begin{equation}
\sigma = \frac{1}{N'} \sum_{i=1}^{N'} \sigma_i.
\end{equation}

When the local stress $\sigma_i$ is lower than the local yield stress $\sS$,
we regard the site $i$ as a sticked site, where
the displacement $u_i$ does not change.
As time $t$ goes on, the local stress $\sigma_i$
given by Eq. \eqref{si_mo} increases.
When the shear stress $\sigma_i$ exceeds $\sS$, 
we assume that the site $i$ slips in the shear direction
and $u_i$ grows to relax the shear stress $\sigma_i$ to the `arrest stress' $\sa$.
The time scale for the local slip may be sufficiently small
so that $Vt$ in Eq. \eqref{si} is regarded as unchanged during a slip.
Thus, the displacement $\delta u_i$ and the local stress drop 
$s_{\rm self}$ due to the slip are respectively 
rewritten as
\begin{eqnarray}
 \label{ui}
\delta u_i & = & - \frac{\sS - \sa}{K+J}, \\
s_{\rm self} & = & - (\sS - \sa),
\end{eqnarray}
which leads to the increase of the local stress at the other sites as
\begin{equation}
s_{\rm oth} = C (\sS - \sa) /N'
\end{equation}
with
\begin{equation}
C = \frac{J}{J + K}.
\end{equation}
Then, 
the stress drop $s$ of the total system is approximately given by
$-(1-C)(\sS - \sa) /N'$.

This increase of the local stress may lead to the slip of 
a site $j \neq i$, and result in a sequential avalanche
with $n$ slips, where the stress drop is given by 
\begin{equation}
s = (1-C)(\sS - \sa)n /N'.
\label{dsn}
\end{equation}

\subsection{Derivation of $\rho(s)$}

As time goes on, the system is expected to reach a statistical steady state.
In this subsection, we derive the probability 
of the stress drop $s$ in the steady state.

\begin{figure}[htbp]
  \includegraphics[width=\linewidth]{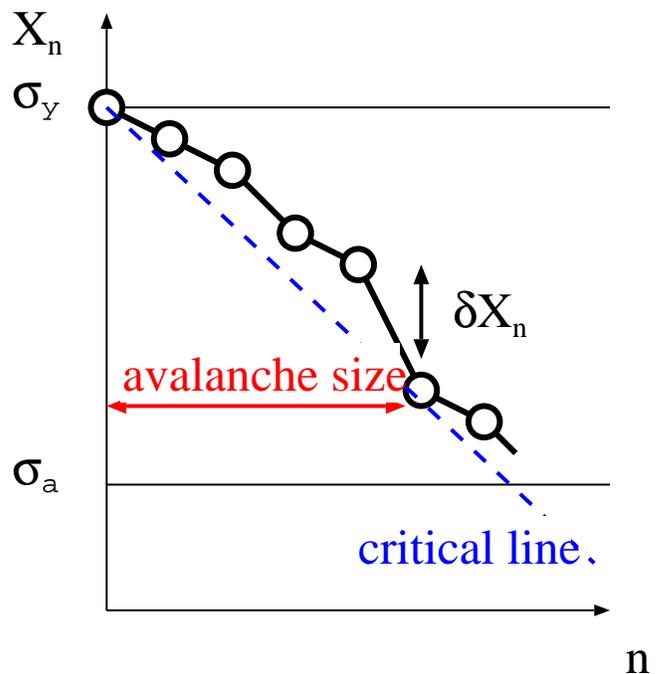}
  \caption{The schematic picture of $X_n$.
  The solid line is the critical line $\sS - n s_{\rm oth}$.}
  \label{X}
\end{figure}

Let us consider the distribution of $\sigma_i$ just before 
the avalanche begins in order to derive the probability of $s$.
Here, we introduce a variable $X_n$ as
\begin{equation}
X_n = \sigma_{i(n+1)},
\end{equation}
where $i(n)$ is the index of the site that has the $n$th largest stress
(see Fig. \ref{X}).
The largest value $X_0$ is $\sS$.
$X_n$ decreases as $n$ increases
with the gap
\begin{equation}
\delta X_n = X_{n-1} - X_{n},
\end{equation}
which is randomly distributed.

Assuming that $\sigma_i$ is 
likely to take any allowable value between $\sa$ and $\sS$,
$X_n$ obeys a Poisson process.
The probability of the intervals divided 
by variables obeying a Poisson process
satisfies an exponential distribution \cite{Feller}.
Therefore, the distribution of $\delta X_n$ is given by
\begin{equation}
\rho_X (\delta X_n) = \frac{N'}{\sS - \sa} e^{-\frac{N'}{\sS - \sa}\delta X_n}.
\label{X:dis}
\end{equation}

When the avalanche starts, the site $i(1)$ slips
and the local stress at other sites increases
by $s_{\rm oth} = C (\sS - \sa) /N'$.
If the local stress at the site $i(2)$ exceeds
$\sS$ because of the increase of the stress, it slips.
This means that
the slip proceeds to the site $i(2)$ if
$X_1$ is larger than $\sS - s_{\rm oth}$.
Similarly, 
the site $i(n+1)$ slips if
$X_n$ is larger than $\sS - n s_{\rm oth}$.
In Fig. \ref{X}, we plot the critical line $\sS - n s_{\rm oth}$.
Therefore,
the size of the avalanche
of the sample shown in this figure
is given by the length of the region where $X_n$ exceeds
the critical line.

\begin{figure}[htbp]
  \includegraphics[width=\linewidth]{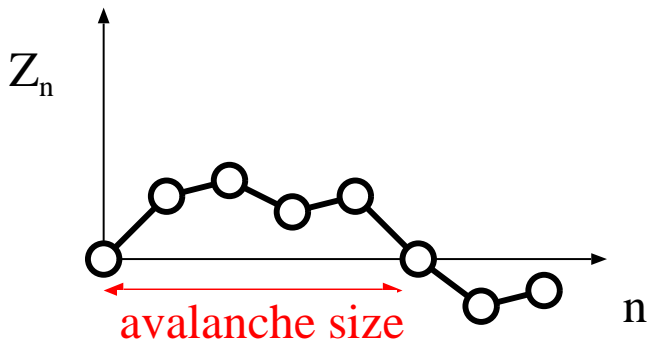}
  \caption{The schematic illustration of $Z_n$.}
  \label{Z}
\end{figure}

In order to obtain the probability distribution of the avalanche size,
we define
\begin{equation}
Z_n = X_n - (\sS - n s_{\rm oth}).
\label{Z:def}
\end{equation}
We plot the schematic illustration of $Z_n$ in Fig. \ref{Z}.
The avalanche size is 
the length of the region where $Z_n$ exceeds $0$.
Since $Z_n$ is considered as a biased random walk,
the avalanche size is calculated as the first passage time
of the biased random walk.

Here, we assume that
$\delta Z_n = Z_n - Z_{n-1}$ obeys a Bernoulli trial
which has $\Delta x$ and $-\Delta x$ with the probability $p$ and $1-p$, respectively.
Then, the average $\mu_Z$ and the variance $V_Z$ are given by
\begin{eqnarray}
\mu_Z & = &  (2p - 1) \Delta x,
\label{muZ:ber}
\end{eqnarray}
\begin{eqnarray}
V_Z & = & 4\Delta x^2 p( 1 - p).
\label{siZ:ber}
\end{eqnarray}

Because $\delta Z_n = Z_n - Z_{n-1}$
is rewritten with Eq. \eqref{Z:def} as
\begin{equation}
\delta Z_n 
= -\delta X_n + s_{\rm oth}
\label{Zn}
\end{equation}
and the probability distribution of $\delta X_n$ satisfies Eq. \eqref{X:dis},
$\mu_Z$ and $\sigma_Z$ are respectively given by
\begin{eqnarray}
\mu_Z & = & -(1-C)\frac{\sS - \sa}{N'}, \label{muz}
\end{eqnarray}
\begin{eqnarray}
V_Z & = & 
 \frac{(\sS - \sa)^2}{N^{'2}}.\label{sz}
\end{eqnarray}
From Eqs. \eqref{muZ:ber}, \eqref{siZ:ber}, \eqref{muz}, and \eqref{sz},
the probability $p$ and the step size $\Delta x$ are respectively given by
\begin{eqnarray}
p & = & \frac{1}{2} \left ( 1 +\frac{C-1}{\sqrt{1+(C-1)^2}} \right ), \\
\Delta x & = & \sqrt{1+(C-1)^2}\frac{\sS - \sa}{N'}.
\label{pa}
\end{eqnarray}

Here, we introduce $\lambda_n$ as 
the probability that $Z_n = \sum_{m=1}^n \delta Z_m$
becomes negative for the first time at the $n$-th step.
As shown in Ref. \cite{Feller},
such probability for the first passage problem
is given by
\begin{eqnarray}
\lambda_{2n-1} & = & 0, \\
\lambda_{2n} & = &
\frac{1}{2p}
\left (
  \begin{array}{c}
  1/2 \\
 n 
  \end{array}
\right )
(-1)^{(n+1)} \{ 4p(1-p) \}^n.
\end{eqnarray}

With the aid of Stirling's formula with Eq. \eqref{pa},
$\lambda_{2n}$ for sufficiently large $n$ is approximately given by
\begin{eqnarray}
\lambda_{2n} =
\frac{1}{4 \pi^{1/2}p}
  \frac{1}{n^{3/2}}e^{-n/n_c},
  \label{lambda:final}
\end{eqnarray}
with $n_c = -1/\log(4p(1 - p)) = 1/\log(1+(C-1)^2)$.

Because the avalanche size $n$ is proportional to the stress drop $s$
as shown in \eqref{dsn},
the probability density $\rho(s)$ 
of the stress drop $s$ 
is thus approximately given by
Eq. \eqref{rhos}.

\begin{figure}
  \includegraphics[width=\linewidth]{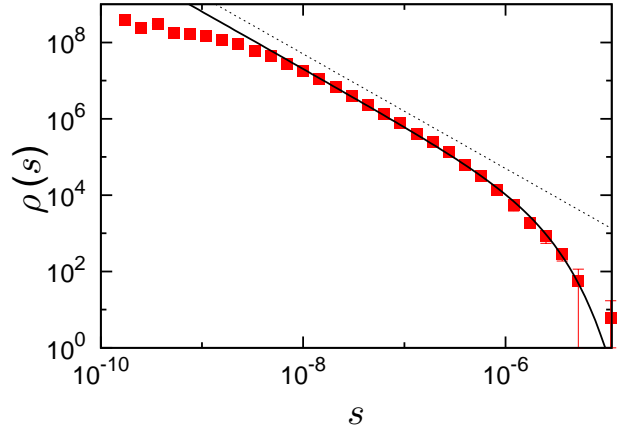}
  \caption{(Color online) 
The probability density $\rho(s)$ 
against $s$ for $\phi_J = 0.6700, \gamma_0 = 10^{-2}$
with $\Delta = 1$.
The dotted and solid lines, respectively, 
represent the power law function with the exponent $3/2$ 
and Eq. \eqref{rhos} with $A = 2.0 \times 10^{-5}, s_{\rm c} = 1.4 \times 10^{-6}$.
}
   \label{hist}
\end{figure}

Figure \ref{hist} is the numerical result of the stress
drop, which well reproduces Eq. \eqref{rhos}
in the region $s > 10^{-8}$,
where
the probability density $\rho(s)$ 
against $s$ for $\phi_J = 0.6700, \gamma_0 = 10^{-2}$
with $\Delta = 1$ is shown.


\begin{thebibliography}{99}
\bibitem{gamma:note}
In order to obtain the state with $\gamma_0$ and $\omega$,
we vary the strain amplitude $\gamma_0$ from $10^0$ to
sequentially decreasing values as $\gamma_0
= 10^0, 10^{-1}, 10^{-2}, 10^{-3}, 10^{-4}, 10^{-5}$.
For each value of $\gamma_0$, 
we sequentially change the values of the frequency $\omega$
as
$\omega \tau= 10^{-1}, 10^{-2}, 10^{-3}, 10^{-4}$.
For a fixed set of $\gamma_0$ and $\omega$,
we measure the stress in a statistical
steady state, where we use the data in the last cycle of
the four cycles of the oscillatory shear.
\end{thebibliography}
\end{document}